\documentclass[apjl]{emulateapj}
\usepackage{natbib}
\slugcomment{Accepted in ApJS, 7 Sep 2013}

\shorttitle{Ring-apodized vortex coronagraph}
\shortauthors{Mawet et al.}

\begin{document}


\title{Ring-apodized vortex coronagraphs for obscured telescopes.\\ 
I. Transmissive ring apodizers.}


\author{D. Mawet\altaffilmark{1}}
\affil{European Southern Observatory, Alonso de Cord\'ova 3107, Vitacura, Santiago, Chile}

\author{L. Pueyo\altaffilmark{2}}
\affil{Space Telescope Science Institute, 3700 San Martin Drive, Baltimore, MD 21218, USA}

\author{A. Carlotti}
\affil{Mechanical \& Aerospace Engineering, Princeton University, Olden street, 08544, Princeton, NJ, USA}

\author{B. Mennesson}
\author{E. Serabyn}
\author{J. K. Wallace}
\affil{Jet Propulsion Laboratory, California Institute of Technology, 4800 Oak Grove Drive, Pasadena, CA 91109, USA}



\altaffiltext{1}{Jet Propulsion Laboratory, California Institute of Technology, 4800 Oak Grove Drive, Pasadena, CA 91109, USA}
\altaffiltext{2}{Department of Physics and Astronomy, Johns Hopkins University, Baltimore, MD, USA}


\begin{abstract}
The vortex coronagraph (VC) is a new generation small inner working angle (IWA) coronagraph currently offered on various 8-meter class ground-based telescopes. On these observing platforms, the current level of performance is not limited by the intrinsic properties of actual vortex devices, but by wavefront control residuals and incoherent background (e.g.~thermal emission of the sky) or the light diffracted by the imprint of the secondary mirror and support structures on the telescope pupil. In the particular case of unfriendly apertures (mainly large central obscuration) when very high contrast is needed (e.g.~direct imaging of older exoplanets with extremely large telescopes or space-based coronagraphs), a simple VC, as most coronagraphs, can not deliver its nominal performance because of the contamination due to the diffraction from the obscured part of the pupil. Here we propose a novel yet simple concept that circumvents this problem. We combine a vortex phase mask in the image plane of a high-contrast instrument with a single pupil-based amplitude ring apodizer, tailor designed to exploit the unique convolution properties of the VC at the Lyot-stop plane. We show that such a ring-apodized vortex coronagraph (RAVC) restores the perfect attenuation property of the VC regardless of the size of the central obscuration, and for any (even) topological charge of the vortex. More importantly the RAVC maintains the IWA and conserves a fairly high throughput, which are signature properties of the VC. 
 \end{abstract}



\section{Introduction}

The main goal of high contrast imaging is to find and, most importantly, characterize extra-solar planetary systems. Indeed, isolating the signal of exoplanets from the glare of their host star enables us to, e.g., measure and constrain their relative orbital motions with precise astrometry, characterize the planetary atmospheres through spectro-photometry, and shed some light on planet-disk interactions \citep[see, for instance,][]{Oppenheimer2009,AbsilMawet2010, Neuhauser2012}. Coronagraphy, which is now a generic term to qualify any techniques used to improve dynamical range in images, promises to be high contrast imaging's sharpest tool, but requires exquisite image quality and stability to perform efficiently. 

The vortex coronagraph \cite[VC,][]{Mawet2005} is one of the most advanced coronagraphs recently brought to operational level \citep{Mawet2011c, Mawet2012}. The VC offers small inner working angle (IWA), potentially down to the diffraction limit ($0.9\lambda/D$), clear 360$^\circ$ off-axis field of view/discovery space, unlimited outer working angle, high throughput, intrinsic and/or induced achromaticity, operational simplicity, and compatibility with the Lyot coronagraph layout. It has also recently demonstrated $\simeq 10^{-9}$ raw contrast levels in the visible on the High Contrast Imaging Testbed at the Jet Propulsion Laboratory \citep[][and, E.~Serabyn et al.~2013, in preparation]{Mawet2012}. It is also at the crux of state of the art high-contrast instruments on various 5-8 $m$ class telescopes. Since it opens a new parameter space at small separations, it has enabled recent  scientific results at Palomar in the H and K bands \citep{Mawet2010a,Mawet2011a,Serabyn2010, Wahl2013}, and at the Very Large Telescope \citep[VLT,][ and O.~Absil et al.~2013, in preparation, and J.~Milli et al.~2013, in preparation]{Mawet2013} in the L' band. It is currently being implemented on SCExAO at Subaru \citep{Martinache2012} and on LMIRCAM at the Large Binocular Telescope \citep[LBT,][]{Skrutskie2010,Esposito2011}. It is also a strong candidate for an exoplanet characterization space-based mission (WFIRST-AFTA, see \citet{Spergel2013a,Spergel2013b}; ACCESS, see \citet{Trauger2010}; and SPICES, see \citet{Boccaletti2012}), for the European-Extremely Large Telescope \citep{Mawet2012}, and the Thirty-Meter Telescope.

\begin{figure*}[!t]
\centerline{\includegraphics[width=12cm]{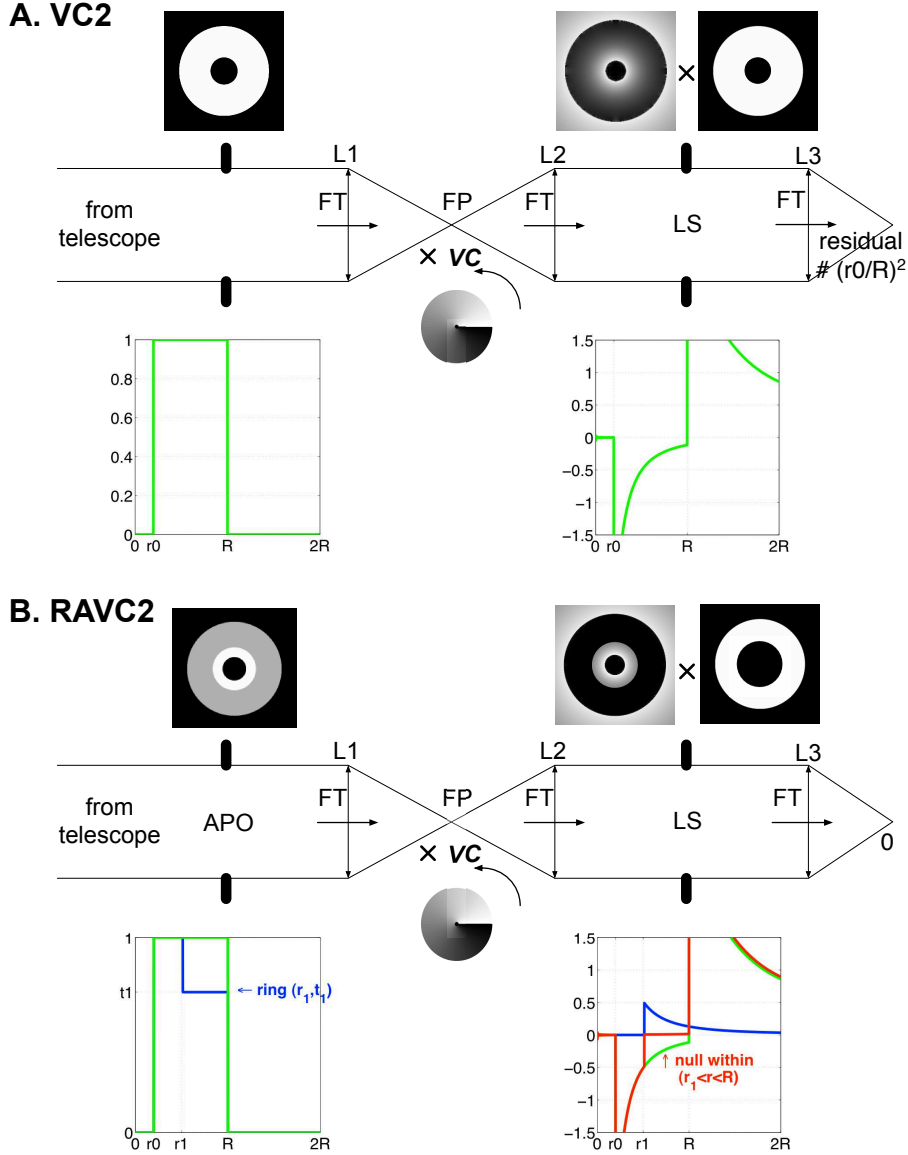}}
\caption{A: classical VC of topological charge 2 with a centrally obscured telescope of radius $R$ ($r_0$ is the radius of the secondary shadow). The residual field interior to the pupil (between $r_0$ and $R$) leads to contrast degradation in the subsequent focal plane image, as the fraction of the total energy remaining inside the pupil is $(r_0/R)^2$, or $0.04$ for a $20\%$ central obscuration. B.: RAVC of topological charge 2. The ring of radius $r_1$ and amplitude transmittance $t_1$ is optimized so that the overlap of the self-similar vortex functions at the Lyot plane issued from the central obscuration (green curve) and the ring (blue curve) perfectly cancel each other between $r_1$ and $R$ (red curve).\label{fig1}}
\end{figure*}

However, as for all other coronagraphs, the VC is sensitive to the aperture geometry, and particularly to secondary obscurations \citep{Mawet2010b,Mawet2011b}. This sensitivity stems from the fact that a Vortex phase ramp in the focal plane of a telescope always diffracts light to the {\em outer regions} of circularly symmetric pupil intensity discontinuities. Thus, as expected, a single vortex will move light outside of the secondary obscuration and support structures, right into the primary pupil image (Fig.~\ref{fig1}, A). The subsequent contrast degradation is proportional to the obscured area $(r_0/R)^2$, with $r_0$ and $R$ being the radii of the central obscuration and primary mirror, respectively \citep{Mawet2010b}.

Recently we proposed a method \citep{Mawet2011b}, based on multiple vortices, that, without sacrificing throughput, reduces this residual light leakage to $(r_0/R)^{2n}$, with $n$ the number of coronagraph stages. This method thus enabled high contrasts to be reached even with an on-axis telescope, but at the cost of increased optical complexity, and for an imperfect result. Here we propose a new simple and elegant solution to this problem that renders the VC completely insensitive to central obscurations with a single VC stage. 

Section~\ref{sec:principle} presents the principle of the ring-apodized vortex coronagraph (RAVC), starting with the charge 2 VC. Section~\ref{sec:RAVC4} develops the charge 4 case, while Section~\ref{sec:generalized} lays out the basis for a generalization to higher order VCs. In Section~\ref{sec:perf}, we discuss the various trade-offs between sensitivity to low-order aberrations, stellar size and throughput. Section~\ref{sec:tech} presents current high-performing technical solutions to manufacture the VC, the apodizer, and to mitigate the diffraction from the support structures, demonstrating the RAVC's high level of technology readiness. Section~\ref{sec:concl} summarizes the concept principles and puts it into the context of future Extremely Large Telescopes (ELT) and space-based missions.

\section{Principle of the RAVC}\label{sec:principle}

The ring-apodized vortex coronagraph (RAVC) is based on the superposition principle and the vortex properties of moving light in and out of circular apertures. Its principle relies on modulating the entrance pupil with one (or a set of) concentric ring(s) of well chosen size(s) and transmittance(s), in order to yield perfect cancellation of on-axis sources at the Lyot stop level. In the following, we show that perfect solutions can be found for any topological charge. We will start with the case of topological charge 2 RAVC, and detail the derivation for charge 4 RAVC in the next section. We finally generalize this concept for arbitrary topological charges in Section~\ref{sec:generalized}.

\subsection{The simple case of charge 2 RAVC}

The effect of a charge $l=2$ vortex phase ramp, $e^{i2\theta}$, applied to the ideal focal plane field (Airy pattern), $\frac{2J_1(k\rho R)}{k\rho R}$, of a filled circular aperture of radius $R$, where k is the wavenumber and $\rho$ is the radial coordinate in the focal plane, has been calculated analytically by various authors \citep{Mawet2005, Jenkins2008, Swartzlander2009, Carlotti2009}. The Fourier transform of focal plane electrical field $e^{i2\theta} \frac{2J_1(k\rho R)}{k\rho R}$ gives the field in the pupil plane downstream from the coronagraph (Lyot-stop plane). Dropping the azimuthal phase term, this transform yields:

\begin{equation}\label{l2e1}
E_{L}(r)=\left \lbrace
\begin{array}{ll}
0 & r<R \\
\left(\frac{R}{r}\right)^2 & r>R
\end{array}
\right.
\end{equation}

Using the superposition principle, a centrally obscured pupil can be seen as the difference between a filled pupil of radius $R$ and a smaller filled pupil of radius $r_0$, yielding a pupil field after the topological charge 2 vortex of \citep{Mawet2011b}

\begin{equation}\label{l2e2}
E_{L}(r)=\left \lbrace
\begin{array}{ll}
0 & r< r_0 \\
-\left(\frac{r_0}{r}\right)^2 & r_0<r<R \\
\left[\left(\frac{R}{r}\right)^2-\left(\frac{r_0}{r}\right)^2\right] & r>R
\end{array}
\right.
\end{equation}

The Lyot stop then blocks everything for $r>R$, so from now on, we will not consider this area anymore in order to focus on the region of interest, i.e.~$r<R$. Indeed, the residual field interior to the pupil (between $r_0$ and $R$) leads to contrast degradation in the subsequent focal plane image, as the fraction of the total energy remaining inside the pupil is $(r_0/R)^2$, or $0.04$ for a $20\%$ central obscuration. 

Consider now that the entrance pupil has an additional ring with $r$ from $r_1$, such that $r_0<r_1<R$, to the outer radius $R$, and characterized by an amplitude transmission coefficient $t_1$ for $r_1<r<R$. Note that the interior of the ring, $r_0<r<r_1$, has a transmission $t_0=1$. Using the same reasoning, we now have within the Lyot plane after the charge 2 vortex

\begin{equation}\label{l2e3}
E_{L}(r)=\left \lbrace
\begin{array}{ll}
0 & r< r_0 \\
-\left(\frac{r_0}{r}\right)^2 & r_0<r<r_1 \\
(1-t_1)\left(\frac{r_1}{r} \right)^2-\left(\frac{r_0}{r}\right)^2 & r_1<r<R 
\end{array}
\right.
\end{equation}

\begin{figure*}[!t]
\centerline{\includegraphics[width=17cm]{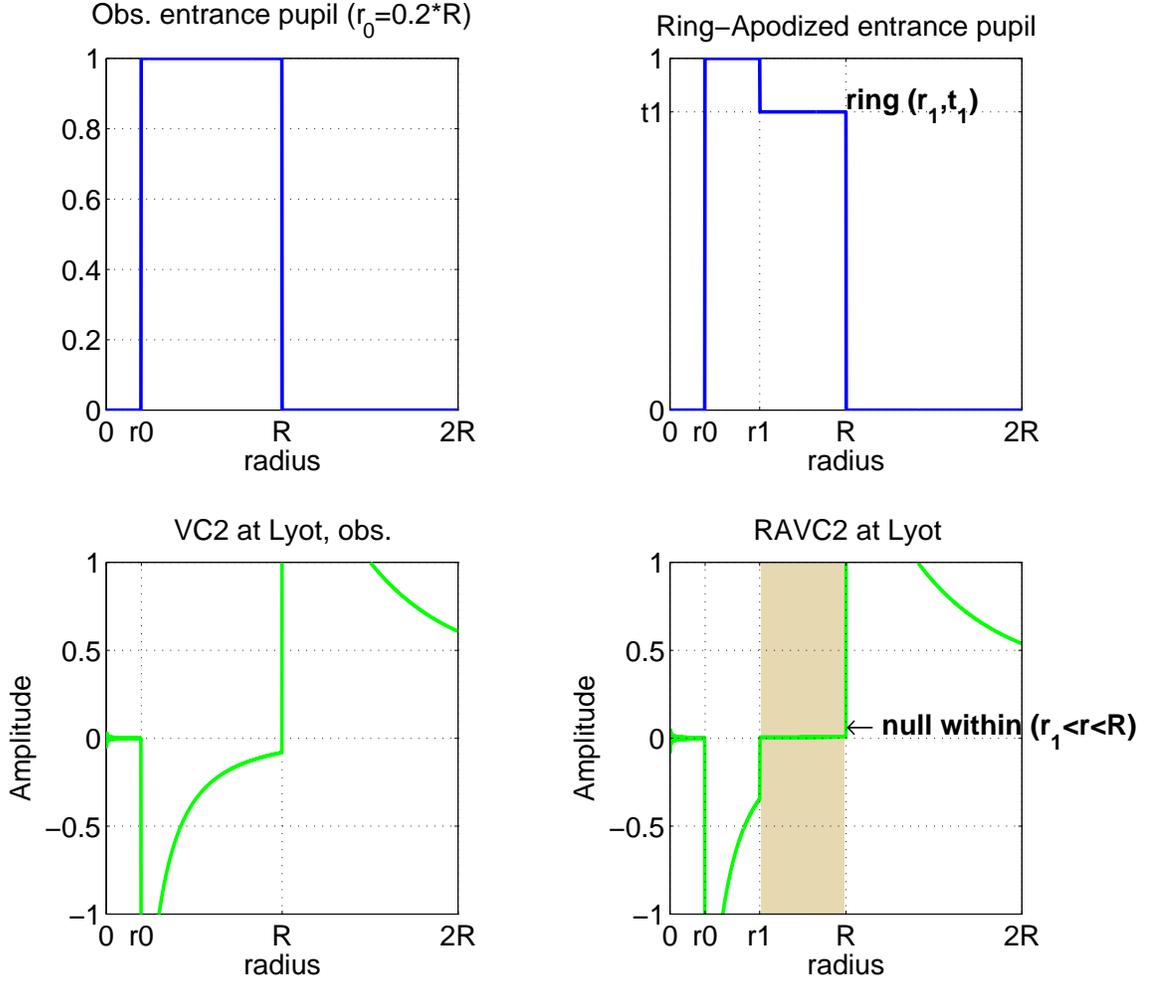}}
\caption{RAVC2. Top left: entrance pupil with central obscuration $r_0=0.2 R$. Top right: ring apodizer of inner radius $r_1$ and amplitude transmittance $t_1$, optimized for maximum throughput. Bottom left: response of the vortex at the Lyot plane showing the contamination from the central obscuration $1/r^2$ vortex function. Bottom right: response of the RAVC at the Lyot plane, showing the perfect null within $r_1<r<R$. \label{fig2}}
\end{figure*}

It appears clearly that the degrees of freedom introduced by the ring apodizer (namely its size $r_1$, and transmittance $t_1$)  provide enough leverage to completely cancel the light within $ r_1<r<R$. Indeed if 

\begin{equation}\label{l2e4}
(1-t_1)=\left(\frac{r_0}{r_1} \right)^2
\end{equation}

then the field in the Lyot plane for $r_1< r <R$ is completely nulled. Fig.~\ref{fig1}, B shows 1-dimensional calculations and 2-dimensional simulations where, as expected, the vortex fields issued from the central obscuration and the ring perfectly balance and cancel each other at the Lyot plane, between $r_1$ and $R$ downstream from the VC (see also Fig.~\ref{fig2}). The single Lyot stop is then designed to block the light for $0<r<r_1$, thus effectively increasing the size of the final central obstruction, and of course for $r>R$.

\subsection{Throughput optimization}

There is a whole set of solutions to Eq.~\ref{l2e4} with $0<t_1<1$ and $r_0<r_1<R$. However, the best solution will maximize the throughput $T$ for a given $r_0$. $T$ is defined as the energy going through the ring $r_1<r<R$, normalized by the energy nominally transmitted by the centrally obscured telescope aperture, or

\begin{equation}\label{l2e5}
T=\frac{t_1^2  \left(1-\left(\frac{r_1}{R} \right)^2\right)}{1-\left(\frac{r_0}{R} \right)^2}
\end{equation}

\begin{figure}[!t]
\begin{center}
\includegraphics[width=8cm]{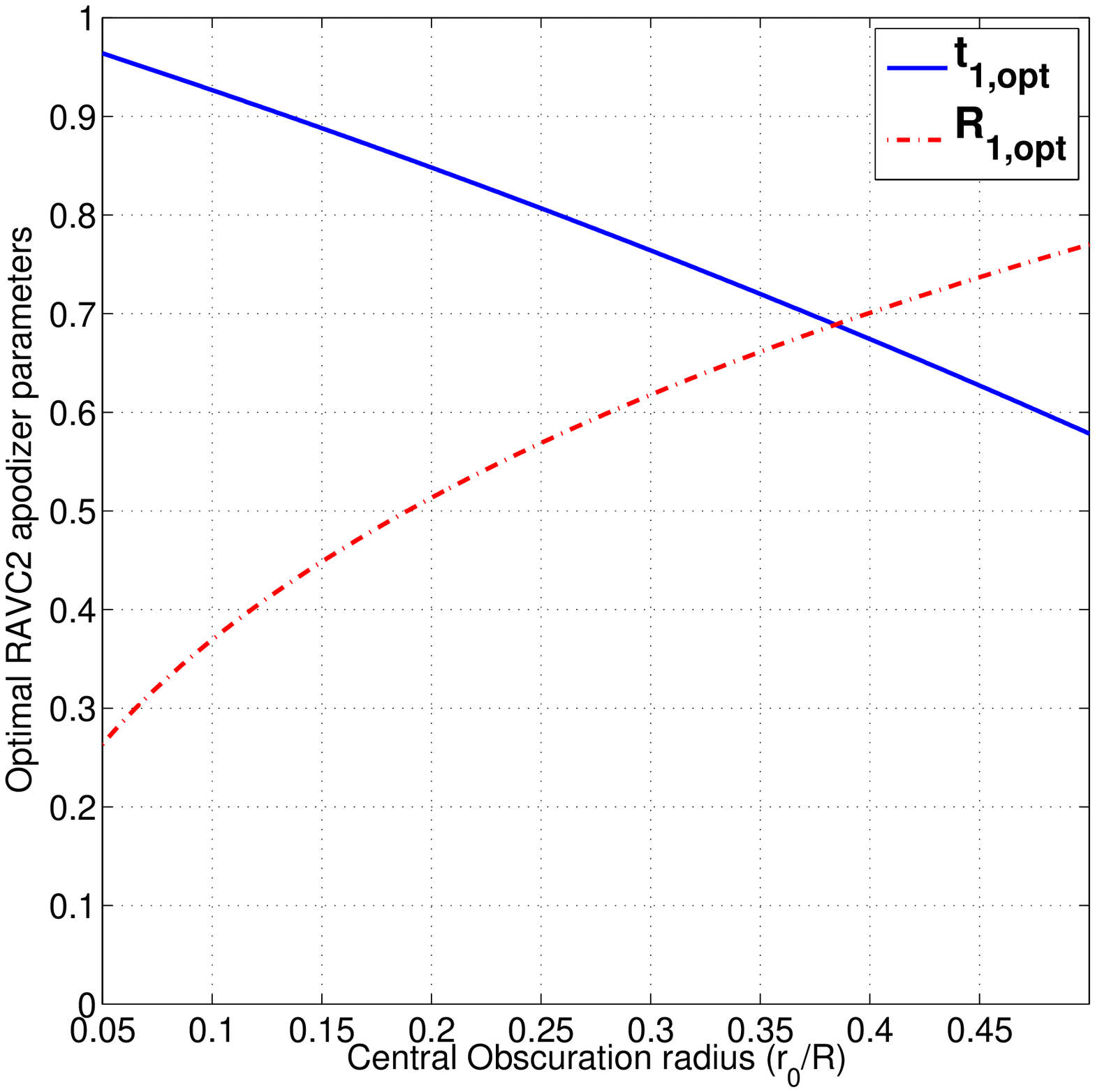}
\caption{Optimal apodizer parameters for a charge 2 ring-apodized vortex coronagraph (RAVC2), $t_{1,opt}$ and $R_{1,opt}$, as a function of $r_0/R$. \label{fig2b}}
\end{center}
\end{figure}

Substituting Eq.~\ref{l2e4} into Eq.~\ref{l2e5}, and differentiating T with respect to $t_1$, we find the optimal ring parameters associated with a charge 2 VC

\begin{equation}\label{l2e6}
\left \lbrace 
\begin{array}{l}
t_{1,opt}=1-\frac{1}{4} \left(R_0^2+R_0 \sqrt{R_0^2+8} \right) \\
R_{1,opt}=\frac{R_0} {\sqrt{1-t_{1,opt}}}
\end{array}
\right.
\end{equation}

where $R_0=r_0/R$, and $R_1=r_1/R$ are the relative radii. Note that $t_{1,opt}$ and $R_{1,opt}$ are functions of $r_0/R$ only (see Fig.~\ref{fig2b}), which is remarkably analogous to the problem associated to designing apodizers for apodized pupil Lyot coronagraphs (APLC) with hard edge focal plane masks \citep{Soummer2005}. Indeed, in both cases, there only exists a unique apodizer configuration that maximizes throughput while yielding a chosen level of starlight extinction. However due to the nature of the VC, this optimal solution turns out to rely on sharp variation of the amplitude profile while the optimal solutions for an APLC are smooth.

\section{Charge 4 RAVC}\label{sec:RAVC4}

The charge 2 RAVC design is simple and the analytical solution very easy to find. The cancellation of the field at the Lyot stop within the outer ring, and the throughput maximization provides two equations that fully and unambiguously characterize the apodizer's two free parameters. The charge 4 case is similar in nature but slightly less trivial.

\subsection{Two rings for perfect cancellation}

As the topological charge of the VC increases, so does the complexity of its response at the Lyot stop plane. Following \citet{Mawet2005} and \citet{Carlotti2009}, for a topological charge 4 vortex, we have

\begin{equation}\label{l4e1}
E_{L}(r)=\left \lbrace
\begin{array}{ll}
0 & r<R \\
2\left(\frac{R}{r}\right)^2-3\left(\frac{R}{r}\right)^4 & r>R
\end{array}
\right.
\end{equation}

The amplitude function after the vortex is now a polynomial of order $-4$, following the topological charge $l$ of the vortex. This function is not self-similar anymore, even though each individual term is. For the sake of simplicity, let us rename this polynomial 

\begin{equation}\label{l4e1b}
V_4(r,R)=2\left(\frac{R}{r}\right)^2-3\left(\frac{R}{r}\right)^4
\end{equation}

A single additional ring will not provide enough leverage to cancel both terms, so we will now consider adding a second ring with $r$ from $r_2$, such that $r_1<r_2<R$, to the radius $R$, and characterized by an amplitude transmission coefficient $t_2$ for $r_2<r<R$. Note that the first ring is now of inner radius $r_1$, such that $r_0<r_1<R$ and outer radius $r_2$, and characterized by an amplitude transmission coefficient $t_1$ for $r_1<r<r_2$. Note that the interior of the first ring, $r_0<r<r_1$, still has a transmission $t_0=1$. 

Using the same reasoning as before, we now have within the Lyot plane after the charge 4 vortex, and this double ring apodizer

\begin{equation}\label{l4e2}
E_{L}(r)=\left \lbrace
\begin{array}{ll}
0 & r< r_0 \\

-V_4(r,r_0) & r_0<r<r_1 \\

(1-t_1)V_4(r,r_1)-V_4(r,r_0) & r_1<r<r_2 \\

(t_1-t_2)V_4(r,r_2)+ & \\
(1-t_1)V_4(r,r_1)-V_4(r,r_0) & r_2<r<R 

\end{array}
\right.
\end{equation}

We are now seeking solutions that perfectly cancel the light within the outer ring $r_2<r<R$, using the four free parameters constraining the ring sizes and transmittances, i.e.~$r_1, r_2$ and $t_1, t_2$.

\begin{equation}\label{l4e3}
(t_1-t_2)V_4(r,r_2)+(1-t_1)V_4(r,r_1)-V_4(r,r_0)=0
\end{equation}

Finding solutions to this under-constrained problem is not straightforward as the $V_4(r,R)$ functions are not self similar. However, by separating the quadratic and fourth order terms, and, since $r>0$, we can rewrite Eq.~\ref{l4e3} as 

\begin{equation}\label{l4e4}
\left \lbrace
\begin{array}{l}
(t_1-t_2)\left(r_2\right)^2+(1-t_1)\left(r_1\right)^2-\left(r_0\right)^2=0\\
(t_1-t_2)\left(r_2\right)^4+(1-t_1)\left(r_1\right)^4-\left(r_0\right)^4=0
\end{array}
\right.
\end{equation}

Eq.~\ref{l4e4} is a system of two equations with 4 unknowns. 

\begin{figure*}[!t]
\centerline{\includegraphics[width=17cm]{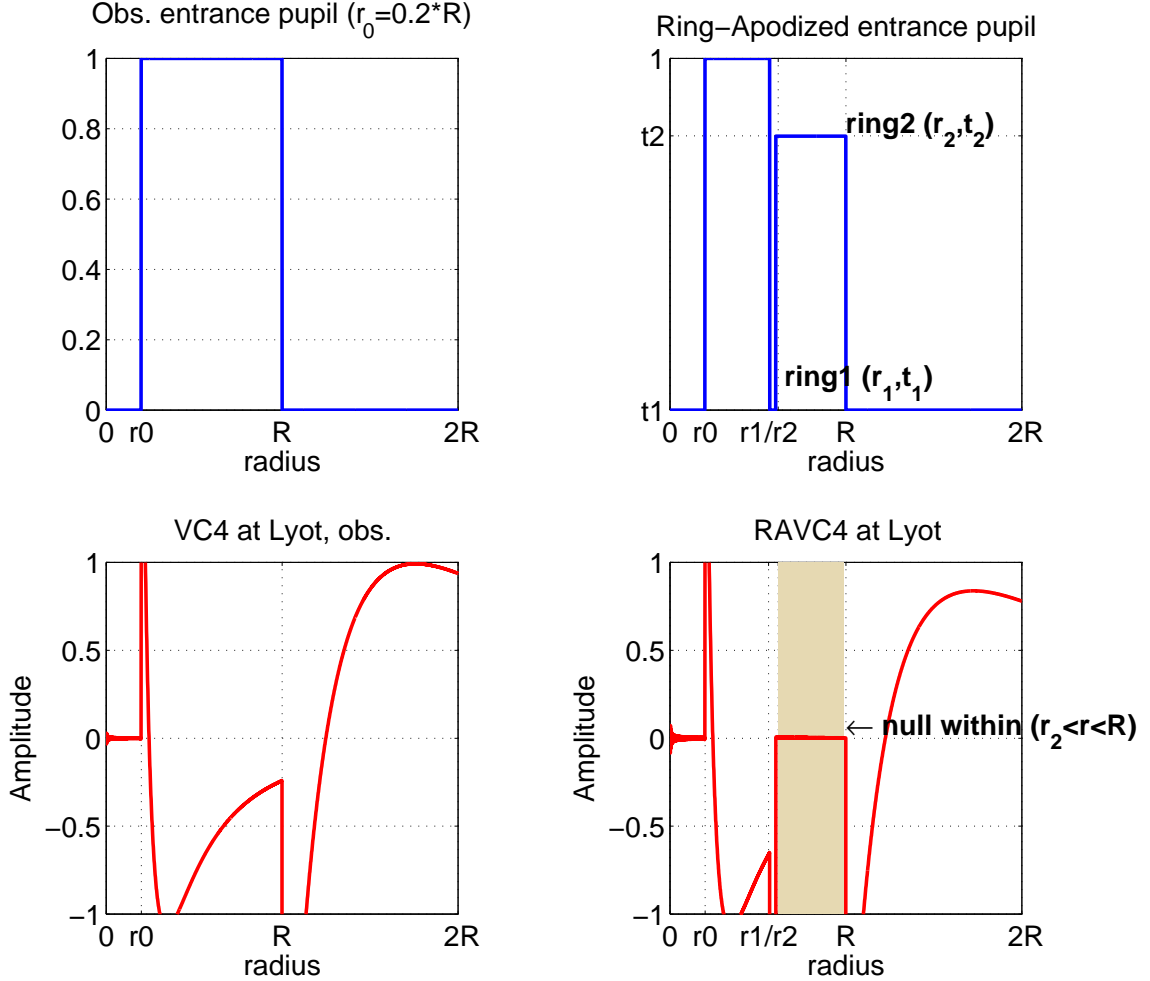}}
\caption{RAVC4. Top left: entrance pupil with central obscuration $r_0=0.2 R$. Top right: ring apodizer, with 2 rings of radius $r_1<r_2$ and amplitude transmittance $0 \le t_1,t_2 \le 1$, optimized for maximum throughput. Bottom left: response of the vortex at the Lyot plane showing the contamination from the central obscuration $V_4(r,R)$ vortex function (see Eq.~\ref{l4e1b}). Bottom right: response of the RAVC4 at the Lyot plane, showing the perfect null within $r_2<r<R$.\label{fig3}}
\end{figure*}

\subsection{Throughput optimization}

In order to better constrain the set of possible solutions, we once again introduce throughput as our figure of merit, but now $T$ is defined as the energy going through the ring $r_2<r<R$.

\begin{equation}\label{l4e5}
T=\frac{t_2^2  \left(1-\left(\frac{r_2}{R} \right)^2\right)}{1-\left(\frac{r_0}{R} \right)^2}
\end{equation}

Closer examination of the throughput expression indicates that optimal solutions are solutions that maximize the outer ring transmittance $t_2$ while keeping $r_2$ as small as possible. Cascading both constraints down to Eq.~\ref{l4e4}, it is easy to derive that such a condition is met for $t_1=0$. Indeed, with $t_1=0$, the modulation terms in $r_1$ and $r_2$ introduced by the rings to balance the central obscuration $r_0$ terms, have maximum weights. Setting $t_1=0$ allows us to simplify the equations greatly, yielding the following charge 4 ring-apodized VC fundamental formulae for optimal throughput

\begin{equation}\label{l4e6}
\left \lbrace
\begin{array}{l}
R_1=\sqrt{\sqrt{R_0^2 (R_0^2+4)}-2R_0^2} \\
R_2=\sqrt{R_1^2+R_0^2}\\
t_2=\frac{R_1^2-R_0^2}{R_1^2+R_0^2}
\end{array}
\right.
\end{equation}

where $R_0=r_0/R$, $R_1=r_1/R$, $R_2=r_2/R$, are the radii relative to the entrance pupil outer radius $R$.  Fig.~\ref{fig3} shows the perfect cancellation of the RAVC4 fields at the Lyot within $r_2<r<R$. One can also explore the entire parameter space $(r_1,r_2,t_1,t_2)$ by solving numerically the following optimization problem:

\begin{eqnarray}\label{l4e7}
&&\mbox{Maximize} \; t_2^2  \left(1-\left(\frac{r_2}{R} \right)^2  \right)\\
&&\mbox{s.t.} 
\left \lbrace
\begin{array}{l}
(t_1-t_2)\left(r_2\right)^2+(1-t_1)\left(r_1\right)^2-\left(r_0\right)^2 =0\\
(t_1-t_2)\left(r_2\right)^4+(1-t_1)\left(r_1\right)^4-\left(r_0\right)^4=0
\end{array}
\right. \nonumber
\end{eqnarray}
This optimization naturally yields solutions for which $t_1 = 0 $, for all sizes of central obscurations. 

\section{Generalization to higher (even) topological charges}\label{sec:generalized}

After detailing the design of charge 2 and 4 RAVCs, for which simple closed form analytical expressions of the apodizer critical dimensioning parameters can be found, we now generalize the concept of the RAVC to higher topological charges.

\subsection{The number of rings is equal to half the charge}

From \citet{Carlotti2009}, we know that for a vortex of topological charge $l$, the vortex function at the Lyot-stop plane downstream from the coronagraph can be written

\begin{equation}\label{le1}
V_{l}(r,R)=i^l\frac{R}{r} Z^1_{l-1}\left( \frac{R}{r} \right) \propto \sum_{j=1}^{l/2} \alpha_j \left(\frac{R}{r} \right)^{2j}
\end{equation}

where $Z^1_{l-1}\left( \frac{R}{r} \right)$ is the radial Zernike polynomial $Z_n^m \left(r \right)$ normalized so that $Z_n^m \left(1 \right) =1$. The real-valued coefficients $\alpha_j$ are computed from the radial Zernike polynomials with, e.g., $\alpha_1 = -1$ for $l=2$, $\alpha_1 = +2$ and $\alpha_2 = -3$ for $l=4$ (see \citet{Carlotti2009} for additional details). Thus the field diffracted in the Lyot plane by the imprint of the central obscuration is always a radial polynomial of order $-l$. As a consequence this polynomial can be perfectly nulled if the coefficient associated with each order is zero. Designing an apodizer with $l/2$ concentric rings, with $r_1<r_2<...<r_{l/2}$ and $t_i>0$, provides enough lever arm to achieve this perfect cancellation. The equations driving the design of the apodizer are then:
\begin{equation}\label{le2}
\sum_{j=1}^{l/2} \left[ (t_{j-1}-t_j)r_j^m \right] - r_0^m=0\  \rm for\  m=2,4,...,l 
\end{equation}
\subsection{Throughput optimization}
For the case of an arbitrary charge vortex it becomes quite challenging to a-priori simplify the problem by setting the transmittance of one or several rings to zero, as we did in the case of a  charge 4 RAVC. However finding the optimal ring design with respect to throughput optimization can be easily carried out by extending the methodology presented in Equation \ref{l4e7}. Indeed The throughput is always a function of the outer ring diameter $r_{l/2}$ and transmittance $t_{l/2}$, as follows
\begin{equation}\label{le3}
T=\frac{t_{l/2}^2  \left(1-\left(\frac{r_{l/2}}{R} \right)^2\right)}{1-\left(\frac{r_0}{R} \right)^2}.
\end{equation}
One can the explore the entire parameter space $(r_1,r_2,...,r_{l/2},t_1,t_2,...,t_{l/2})$ by solving the following optimization problem:
\begin{eqnarray}\label{le4}
&&\mbox{Maximize} \; t_{l/2}^2  \left(1-\left(\frac{r_{l/2}}{R} \right)^2\right)\\
&&\mbox{s.t.} 
\sum_{j=1}^{l/2} \left[ (t_{j-1}-t_j)r_j^m \right] - r_0^m=0\  \rm for\  m=2,4,...,l  \nonumber
\end{eqnarray}
Solving this system of equations for the $r_j$ and $t_j$ is non trivial for higher topological charges, and requires numerical optimization methods. We have verified the existence of solutions for charges up to $l=8$.

\section{Performances}\label{sec:perf}

Here we discuss the performance of the RAVC family in terms of contrast, (off-axis) throughput, and inner working angle. For perfect optics and perfect VCs of various (even) topological charges, there exists RAVC solutions providing infinite contrast whatever the central obscuration. Throughput is a decreasing function of the topological charge and central obscuration size (see Eq.~\ref{le3}). Indeed, throughput is always a function of the outer ring area, which gets smaller when the charge increases (more rings necessary) and of course when the central obscuration gets larger.
\begin{figure}[!t]
\begin{center}
\includegraphics[width=8cm]{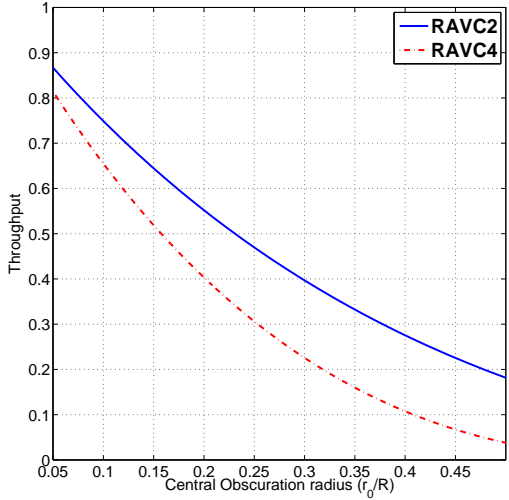}
\caption{Theoretical maximum throughput of the RAVC2 and RAVC4 with transmissive ring apodizers for various obscuration relative diameters $r_0/R$. The throughput decreases with the topological charge and central obscuration. \label{fig4}}
\end{center}
\end{figure}

Higher topological charges $l$, which trade off inner working angle (e.g.~IWA$_{l=2}=0.9\lambda/D$, IWA$_{l=4}=1.75\lambda/D$), are desired when the telescope size increases (to mitigate the stellar size effect) or when sensitivity to low-order aberrations becomes the limiting factor \citep{Mawet2010b}. Indeed, \citet{Jenkins2008} showed that the sensitivity of the VC to pointing offsets $\theta$, in units of $\lambda/D$, is proportional to $\theta^l$, with $\theta << 1$ (the same laws apply to the sensitivity to stellar size, which can be seen as an incoherent sum of pointing offsets).

\begin{figure*}[!t]
\begin{center}
\includegraphics[width=8cm]{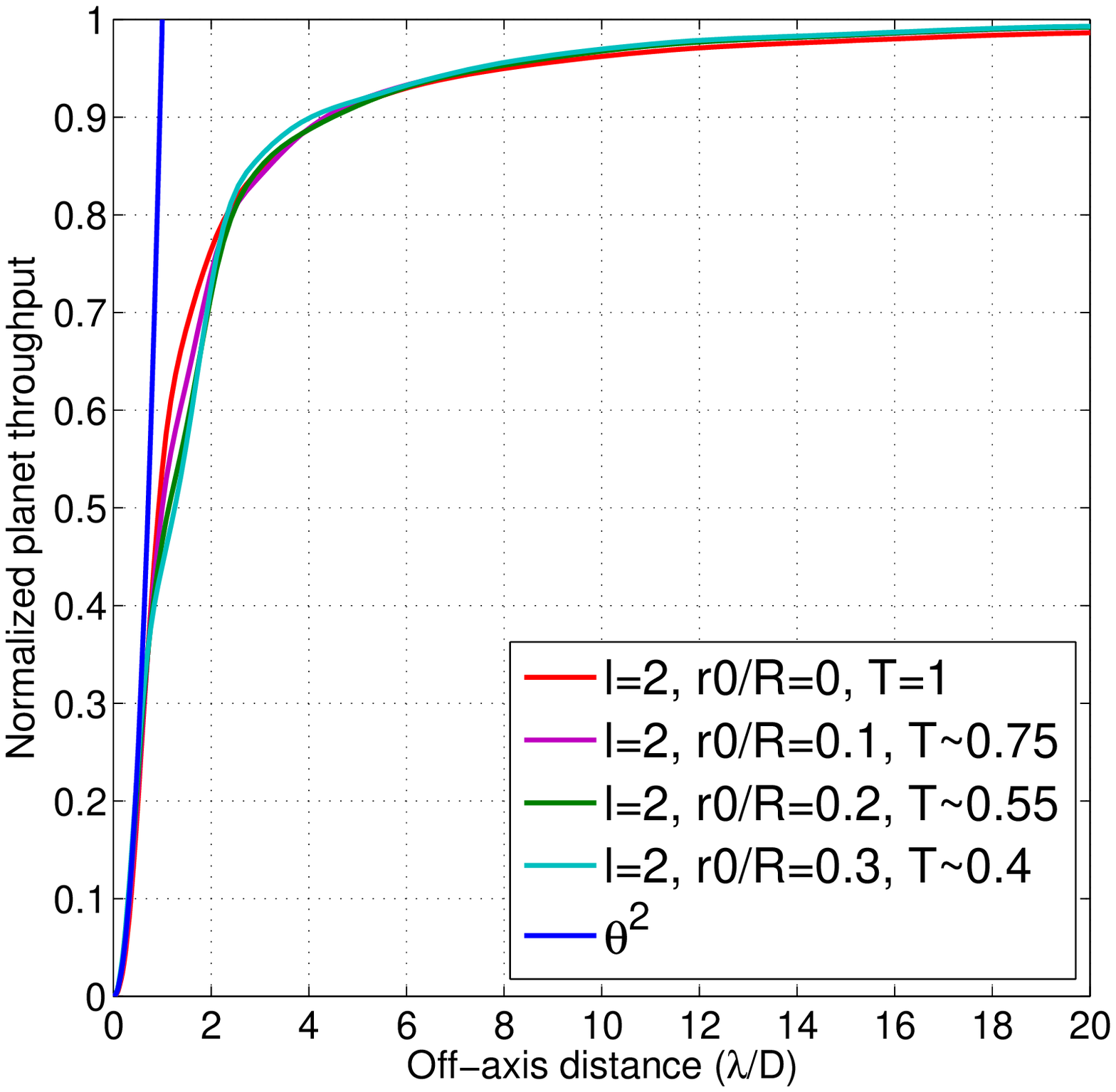}
\hspace{1cm}
\includegraphics[width=8cm]{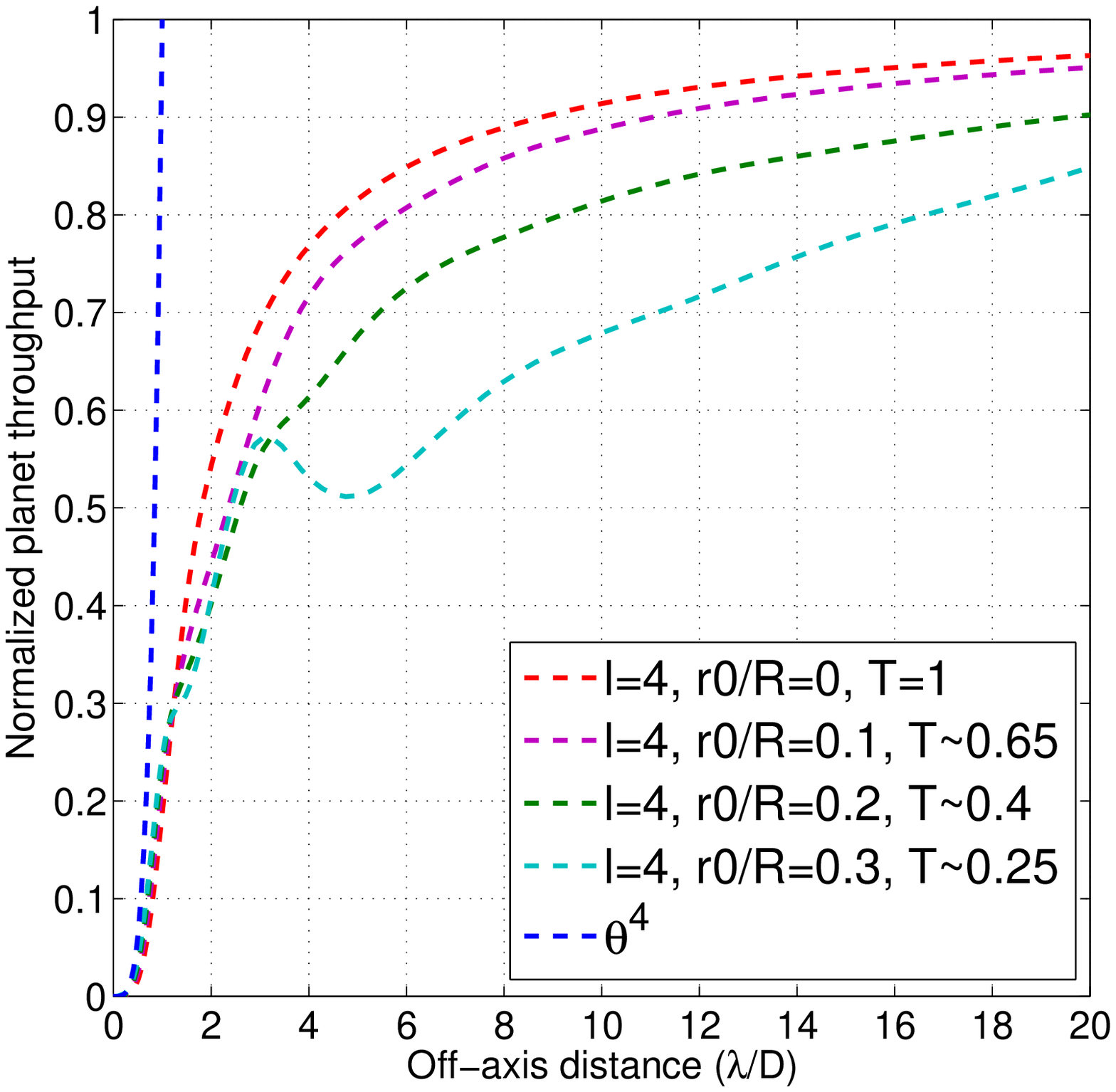}
\caption{Left: normalized off-axis companion throughput for the RAVC2 (charge $l=2$) as a function of angular separation in $\lambda/D$ units. Right: normalized off-axis companion throughput for the RAVC4 (charge $l=4$) as a function of angular separation in $\lambda/D$ units. The different curves are for different obscuration ratios. We over-plotted the $\theta^l$ function, with the pointing offsets $\theta$ in units of $\lambda/D$, representative of the VC sensitivity to low-order aberrations (here tip-tilt) for $\theta << 1$.  \label{fig5}}
\end{center}
\end{figure*}

Fig.~\ref{fig4} presents a throughput curve for the RAVC2 and RAVC4, as a function of central obscuration size. For a $10\%$ central obscuration, the RAVC2 throughput is $\simeq 75\%$ and $\simeq 65\%$ for the RAVC4. For $20\%$, the RAVC2 throughput is $\simeq 55\%$ and $\simeq 40\%$ for the RAVC4. Note that the IWA of the VC, classically defined as the $50\%$ off-axis throughput point (relative to the maximum), is not affected by the apodizer in the topological charge 2 case (see Fig.~\ref{fig5}, left), but marginally affected for the charge 4 case (see Fig.~\ref{fig5}, right), especially as the size of the central obscuration increases.

The RAVC solution is thus a good compromise between the numerically-optimized apodizer masks presented in Carlotti et al.~2013 (in preparation, see also Section~\ref{subsec:strutsapo}) as it has comparable throughput but with a full search area, and the Phase-Induced Amplitude Apodization Complex Mask Coronagraph \citep[PIAACMC,][]{Guyon2013}, which involves more complicated optics. 

\section{Technical feasibility}\label{sec:tech}

In this section we discuss the technical feasibility of the RAVC, from the current technology readiness of the VC to the ring apodizer manufacturability and the optical layout of the concept, including three practical solutions to mitigate the diffraction from the secondary support structures.

\subsection{Vortex mask manufacturing}

The vector vortex coronagraph \citep[VVC,][]{Mawet2005,Mawet2009} is one possible and easy route to manufacture VCs\footnote{Noteworthy progress was recently made in the scalar vortex technology, here using computer generated holograms, see \citet{Errmann2013}.}. It advantageously makes use of the geometrical or Pancharatnam-Berry phase, which is achromatic by nature. The VVC is based on a space-variant halfwave plates (HWP), circularly symmetric in the charge 2 case. Manufacturing the VVC thus requires manipulating the polarization vector in a space-variant manner, i.e.~it needs to be significantly modulated across spatial scales of less than a mm, with precisions of a few microns and fractions of a degree. Three technological approaches are currently used to manufacture the VVC \citep{Mawet2012}: liquid crystal polymers \citep{Mawet2009}, subwavelength gratings \citep{Mawet2005,Delacroix2013}, and photonic crystals \citep{Murakami2010a,Murakami2013}. Each one of these technological choices has advantages and drawbacks, enumerated in \citet{Mawet2012}, and practical vortex devices that have already provided very high contrast with unobscured apertures are already available (see Table~\ref{tab:1}). Thus we now turn to the manufacture of the new component needed, i.e., the ring apodizerÓ.

\begin{deluxetable}{lllll}

\tabletypesize{\scriptsize}

\tablecaption{Characteristics of the three main technologies currently being used to render the VVC (LCP=liquid crystal polymer, SG=subwavelength gratings, PC=photonic crystals). $l$ is the topological charge of the vortex. NIR stands for near-infrared. MIR stands for mid-infrared. ``Cent. def.'' is the size of the defect at the center of the VVC. \label{tab:1}} 


\tablehead{
\colhead{Tech.} &\colhead{$\lambda$} &\colhead{$l$} &\colhead{Cent. def.} &\colhead{Raw Contrast}}

\startdata
LCP &VIS-NIR &2-4 &$<20\mu$m (\tablenotemark{a}) &$\simeq 10^{-9}$ @ 785 nm \\
 & & &$<5\mu$m (\tablenotemark{b}) &$\simeq 2\, 10^{-8}$ 10\% BW  	\\
 & & & &$\simeq 4\, 10^{-8}$ 20\%  BW  		\\
PC &VIS(-NIR) &2 &$<1\mu$m (\tablenotemark{c}) &$\simeq 10^{-8}$ @ 785 nm 	\\
SG &(NIR-)MIR  &2 &$<5\mu$m (\tablenotemark{d}) &$\simeq10^{-5}$ @ 4 $\mu$m (\tablenotemark{e}) 
\enddata
\tablenotetext{a}{Manuf.~by JDSU \citep{Mawet2009,Mawet2012}, see also Serabyn et al.~2013, in preparation.}
\tablenotetext{b}{Manuf.~by BeamCo \citep{Nersisyan2013}.}
\tablenotetext{c}{Manuf.~by Photonic Lattice Inc.~\citep{Murakami2010a, Murakami2013}.}
\tablenotetext{d}{Manuf.~by Uppsala University~\citep{Delacroix2013}}
\tablenotetext{e}{Without wavefront control.}
\end{deluxetable} 

\subsection{Apodizing Mask manufacturing}\label{subsec:apomanuf}

Given the extreme simplicity of the ring apodized masks, and their discrete levels of transmittance, no difficulty is foreseen in this area. The manufacturing of the ring apodizer pupil mask should thus be straightforward and one can envision using either microdot or optical coating technologies.

The microdot technology uses a halftone-dot process, where the relative density of a binary array of pixels (transmission of 0 or 1 at the micron level) is calculated to obtain the required local transmission (here uniform within the rings). The manufacturing of current apodized pupil Lyot coronagraphs \citep{Soummer2005, Soummer2011} for SPHERE \citep{Kasper2012} and GPI \citep{Macintosh2008} uses the microdot technology which is well mastered \citep{Martinez2009a,Martinez2009b}. Note the band-limited coronagraphs of NIRCAM soon to fly aboard JWST have also made use of a similar technique \citep{Krist2009}. The demonstrated advantages of microdot apodizers are numerous: 1\%-level accuracy on the transmission profile, achromatic in phase and amplitude, compatibility with a wide range of substrate material, and with conventional AR coating. Spatial phase distortion are in principle absent \citep{Martinez2009a,Martinez2009b}, but careful control will be necessary for the RAVC. Indeed, the perfect superposition of the fields originating from the central obscuration and the ring(s) requires a uniform phase across the apodizer area.

Another potential technique could make use of optical coatings. \citet{Trauger2012} developed a successful method to induce a quasi-achromatic spatially variable optical density with a combination of a deposited metal together with a dielectric to cancel the induced phase shift. This technology has been used to manufacture the band-limited coronagraph currently holder of the contrast world record \citep{Trauger2011}, and spatial transitions of the order of a few microns should be possible.

\subsection{RAVC Layout}

The RAVC layout is quite simple and only requires the insertion of the apodizer at a pupil plane upstream of the coronagraph (see Fig.~\ref{fig1}, B). Provided that the pupil plane can be shared with a potential deformable mirror, or that the deformable mirror can be slightly out of the pupil plane, no additional stage is required \citep{Mawet2011b}. Such a configuration allows implementing the RAVC on existing ground-based instruments with little additional effort as wheels with pupil masks are most of the time available.

\subsection{Strut mitigation with ACAD}

The analytical solutions presented above only deals with a central on-axis obscuration. Large telescope apertures rarely resemble uniform disks, or annuli. Besides the central obscuration, they usually feature opaque areas produced for example by their support structures or gaps between main mirror segments. Following the superposition principle, such opaque areas diffract light in the same way but with opposite phase. The resulting PSF structure produced by opaque areas can be detrimental to high-contrast imaging, and the total scattered flux is proportional to the size of the obscured area.

Secondary mirror spiders produce extended spike-shaped features, and the net-like gap structure of a segmented mirror produces a regular speckle pattern with a pitch that is inversely proportional to the segment pitch. An efficient diffraction control system has to take aperture irregularities into account. Conventional Apodizers have been calculated for irregular apertures \citep{Carlotti2011}, and are now optimized to deal with phase mask coronagraphs \citep[see Sect.~\ref{subsec:strutsapo} below, and][]{Carlotti2013}.

An interesting alternative to classical apodization techniques is the upfront correction of aperture irregularities by optical remapping in the geometric and thus achromatic regime. While PIAA can remove central obscurations, \citet{Pueyo2012} presented a method (ACAD, Active Correction of Aperture Discontinuities) to derive mirror shapes suitable to remove the narrow structures introduced by spiders, gaps and maybe even missing segments. Because the required mirror deformations are small and of the order of a few microns, deformable mirrors (DMs) could be used for this purpose.

Even without apodization or remapping, PSF structures produced by gaps and spiders are typically less localized and less affected by the coronagraphic mask. Therefore, they show up mostly as bright structures of the original geometry in the Lyot plane of an efficient coronagraph and can be masked to a large extent by a suitable irregularly shaped Lyot stop. For larger separations the fraction of the field of view spoiled by spiders and gaps may be sufficiently small to ignore.

\subsection{Strut mitigation by apodizer optimization}\label{subsec:strutsapo}

If two DMs are available, ACAD can be used to mitigate the diffraction effects due to the struts. However, if there is only one DM available, or no DMs at all, this task can be given to a different type of apodizer, specifically computed to take these additional diffraction effects into account. Following an idea first presented in \citet{Carlotti2013}, and then applied to the case of the four-quadrant phase mask, the 2D transmission of amplitude apodizers can be maximized in a numerical optimization problem, where constraints are set on the extremum values of the electric field in the Lyot plane.

An upcoming paper (A.~Carlotti et al.~2013b, in preparation) presents charge-2 and charge-4 VC apodizers designed for several on-axis telescopes with 10-30\% central obscurations and orthogonal spiders. Interestingly, the overall morphology of these numerically optimized solutions converges to the analytical RAVC design for the ideal strut-less pupil, and departs from it only  around the struts, where additional local apodization features are necessary. While slightly smaller, the transmissions of these apodized coronagraphs are also comparable to the transmissions of the RAVC2 and RAVC4. This is mostly due to the presence of the secondary supports in the pupil, but also partially explained by the finite radius imposed to the vortex phase mask (currently limited to 32-64 $\lambda/D$ because of the complexity of the computations). As other 2D optimal apodizers, these masks have binary transmissions, and can thus be manufactured the same way the shaped pupil coronagraph \citep[SPC,][]{Kasdin2007} was manufactured, i.e.~using Deep Reactive Ion Etching (DRIE) of a Silicon-On-Insulator (SOI) wafer. The fabrication of binary apodizers can also make use of the microdots and coating technologies, as discussed in Section~\ref{subsec:apomanuf}. 

Another straightforward strut mitigation technique is the spider removal plate (SRP) which removes the strut footprint by translating the clear and contiguous parts of the pupil inwards with tilted plane-parallel plates \citep{Lozi2009}. This solution is however less ideal for very high contrast applications since it introduces a thick prismatic optical element in the beam upstream of the coronagraph, and with it, its share of chromatic optical aberrations.

\section{Conclusions : a game-changing concept}\label{sec:concl}

The ring-apodized vortex coronagraph (RAVC) is a game-changing concept. It unambiguously solves the last hurdle that the VC faced, namely its sensitivity to central obscuration. Contrary to previous solutions which were relying on multi-stage approaches \citep{Mawet2011b} or complex numerically optimized apodization solutions \citep{Carlotti2013}, the RAVC is a single stage approach with extremely simple apodizer designs. The simplicity of the RAVC concept enables fast track implementation.

The concept is particularly relevant to future extreme adaptive optics instruments for Extremely Large Telescopes (ELT) and coronagraphic space missions employing on-axis telescopes, where central obscuration and the desired use of topological charge 4 VC are additional constraints that the RAVC family solves pragmatically. With a more limited aperture in space, throughput loss might be an issue. However, a forthcoming paper (Pueyo et al.~2013, in preparation) will extend the RAVC concept to lossless apodization techniques, which should mitigate this problem as well.

\acknowledgments

This work was carried out at the European Southern Observatory (ESO) site of Vitacura (Santiago, Chile), and at the Jet Propulsion Laboratory, California Institute of Technology, under contract with National Aeronautics and Space Administration (NASA). This material is partially based upon work supported by NASA under Grant NNX12AG05G issued through the Astrophysics Research and Analysis (APRA) program . This work was performed in part under contract with the California Institute of Technology funded by NASA through the Sagan Fellowship Program executed by the NASA Exoplanet Science Institute.

\bibliography{aporing_rev_v3_arxiv}

\end{document}